\documentstyle[preprint,eqsecnum,aps]{revtex}
\begin{document}
\title{The Schwinger $SU(3)$ construction - I: 
Multiplicity problem and relation to induced representations}
\author{S. Chaturvedi\thanks{email: scsp@uohyd.ernet.in}}
\address{ School  of Physics, University of Hyderabad, Hyderabad 500046,
India}
\author{N. Mukunda\thanks{email: nmukunda@cts.iisc.ernet.in} \thanks{Honorary 
Professor, Jawaharlal Nehru Centre for
Advanced Scientific Research, Jakkur, Bangalore 560064}}
\address{Centre for Theoretical Studies, 
Indian Institute of Science,
Bangalore 560012, India}
%\date{\today}
\maketitle
%\draft{16.8.2000}
\begin{abstract}
The Schwinger oscillator operator representation of $SU(3)$
is analysed with particular reference to the problem of 
multiplicity of irreducible representations.  It is
shown that with the use of an $Sp(2,R)$ unitary 
representation commuting with the $SU(3)$ representation, 
the infinity of occurrences of each $SU(3)$ irreducible 
representation can be handled in complete detail.  A natural 
`generating representation' for $SU(3)$, containing each
irreducible representation exactly once, is identified
within a subspace of the Schwinger construction; and
this is shown to be equivalent to an induced representation
of $SU(3)$.
\end{abstract}

\section{Introduction}
The well known Schwinger representation of the Lie algebra of $SU(2)$
\cite{1}, 
constructed using the annihilation and creation operators of two independent 
quantum mechanical harmonic oscillators, has played an important role in many 
widely differing contexts. Within the quantum theory of angular momentum it 
has made the calculation of various quantities somewhat easier than by other 
methods. Beyond this, it has been very effectively exploited in the physics 
of strongly correlated systems\cite{2},  in quantum optics of two mode 
radiation fields\cite{3}, and in the study of certain classes of partially 
coherent optical beams\cite{4}, namely to obtain the coherent mode 
decomposition of anisotropic Gaussian Schell model beams. 
It has also been used in a recent investigation of the Pauli spin- statistics 
theorem\cite{5}. 

Bargmann has presented an entire function Hilbert space analogue of the 
Schwinger construction, which is extremely elegant and possesses special 
merits of its own\cite{6}. This may be viewed as a counterpart to the 
Fock space description of quantum mechanical oscillator systems. 

Certain specially attractive features of the Schwinger $SU(2)$ construction 
should be mentioned. It leads upon 
exponentiation to a unitary representation 
(UR) of $SU(2)$ in which each unitary irreducible representation (UIR), 
labelled as usual by the spin quantum number $j$ with possible values 
$0,1/2,1,\cdots, $ appears exactly once. 
In other words, it is complete in the 
sense that no UIR of $SU(2)$ is missed, and also economical in the sense of 
being multiplicity free. Thus, reflecting these two features, it may be 
regarded as a `Generating Representation' for $SU(2)$, a concept that has 
been effectively used in understanding the structures of various kinds of 
Clebsch-Gordan series for UIR's of the non compact group $SU(1,1)\cite{7}$. 
In addition of course the use of boson operator methods makes many operator 
and state vector calculations relatively easy to carry out. 

It is of considerable interest to extend the Schwinger construction to other 
compact Lie groups, the next natural case after $SU(2)$ being $SU(3)$. The 
aims behind any such attempt would be to preserve the simplicity of the boson 
calculus, to cover all UIR's of the concerned group, and to do it in a 
multiplicity free manner.

The case of $SU(3)$ has been studied by several authors since the 
work of Moshinsky\cite{8}. The aim of the present paper is somewhat 
different from previous studies, being 
motivated by the particular points of view mentioned above. In particular our 
aim is to see to what extent the attractive features of the $SU(2)$ 
construction survive when we consider $SU(3)$, and which ones have to be 
given up.

 A brief overview  of this paper is as follows. In Section II we collect 
together some relevant facts regarding unitary representations of compact Lie 
groups with special attention to $SU(3)$. In particular, we highlight the 
fact that the theory of induced representations leads to a unitary 
representation of $SU(3)$ which has all the properties becoming of a 
`Generating Representation' of $SU(3)$ in that it contains all the UIR's of 
$SU(3)$ exactly once each. The Hilbert space carrying this unitary 
representation turns out to be the Hilbert space of functions on unit sphere 
in $C^3$. In Section III, we turn to the Schwinger oscillator construction 
for $SU(3)$ and show that a naive extension of the Schwinger 
$SU(2)$-construction making use of six oscillators leads to a very 
`fat' UR of $SU(3)$ containing each UIR of $SU(3)$ infinitely many times. 
We then show how the group $Sp(2,R)$ enables us to completely handle this 
multiplicity and also neatly isolate from  this rather large space a 
subspace carrying  a UR of $SU(3)$ of a `Generating Representation' type. 
At this stage, we have two 'Generating Representations' of $SU(3)$, one based 
on the Hilbert space of functions on a unit sphere in $C^3$ and the other 
based on the Fock space of six oscillators, and a natural question to ask 
is how the two are related. To this end, in Section IV, we make use of the 
Bargmann representation, to transcribe the Fock space description into a 
description based on a Hilbert space of square integrable functions in six 
complex variables satisfying certain conditions. This transcription enables 
us to establish an equivalence map between the Hilbert spaces supporting the 
two incarnations of the 'Generating Representation' for $SU(3)$, details of 
which are given in Sections V and VI. Section VII contains concluding remarks 
and further outlook and an appendix gives the details of the construction 
of $SU(3) \times Sp(2,R)$ basis states. 
    
\section{Unitary representations of compact Lie groups, the $SU(3)$ case}

It is useful to first recall some basic facts concerning the representation 
theory of any compact simple Lie group $G$. The basic building blocks are the 
UIR's of $G$. Each UIR carries certain identifying labels (eigenvalues of 
Casimir operators), such as $j$ for $SU(2)$. It is of a characteristic 
dimension, such as $2j+1$ for $SU(2)$. In addition, we may set up some 
convenient orthonormal basis in the space of the UIR, as simultaneous 
eigenvectors of some complete commuting set of hermitian operators. The 
eigenvalue sets labelling the basis vectors are generalisations of the single 
magnetic quantum number $m$ for $SU(2)$.

A general UR of $G$ is reducible into UIR's, each occurring with some 
multiplicity. Thus the UR as a whole is in principle completely determined 
upto equivalence by these multiplicities. However certain UR's have special 
significance, reflecting the way they are constructed, and so deserve special 
attention. We consider two cases - the regular representation, and 
representations induced from various Lie subgroups of $G$.  

The Hilbert space carrying the regular representation of $G$ is the space 
$L^2(G)$ of all complex square integrable functions on $G$, the integration 
being with respect to the (left and right) translation invariant volume 
element on $G$. On this space there are in fact two (mutually commuting) 
regular representations of $G$, the left and the right regular representations. 
Upon reduction into UIR's each of these contains every UIR of $G$ without 
exception, the multiplicity of occurrence of a particular UIR is just its 
dimension. Thus the regular representations possess the completeness property 
of the Schwinger $SU(2)$ construction, but not its economy. 

Next we look at the family of induced UR's of $G$\cite{9}. Let $H$ be some Lie 
subgroup of $G$, and let $D(h), h\in H$, be the operators of a UIR of $H$ on 
some Hilbert space ${\cal V}$. Then a certain unique UR of $G$, with operators 
${\cal D}_{H}^{({\rm ind}, D)}(g)$ for $g\in G$, can be constructed. 
As the labels 
indicate, this UR is induced from the UIR $D(\cdot)$ of $H$. The Hilbert 
space 
${\cal H}_{H}^{({\rm ind},D)}$ of this UR consists of functions on $G$ with 
values 
in ${\cal V}$ obeying a covariance condition and having finite norm :
\begin{eqnarray}
\psi \in {\cal H}_{H}^{({\rm ind}, D)}~&:&~\psi(g)\in {\cal V}, g\in G \nonumber\\
\psi(gh)&=& D(h^{-1})\psi(g), h\in H \\
||\psi||^2 &=&\int_{G} dg (\psi(g),\psi(g))_{\cal V} < \infty.\nonumber 
\end{eqnarray}
Here $dg$ is the  (suitably normalised) invariant volume element on $G$, and 
the 
integrand is the squared norm of $\psi(g) \in {\cal V}$. The covariance 
condition means that $\psi(g)$ is essentially a function on the coset space 
$G/H$, in the sense that the `values' of $\psi(g)$ all over a coset are 
determined by its `value' at any one representative point. 
Correspondingly due 
to unitarity of $D(h)$, $(\psi(g),\psi(g))_{\cal V}$ is constant over each 
coset; so the expression for $||\psi||^2$ can be simplified and expressed in terms of a $G$-invariant volume element on $G/H$. The action of  
${\cal D}_{H}^{({\rm  ind}, D)}(g)$ on $\psi$ is then given by 
\begin{eqnarray}
g\in G &:& {\cal D}_{H}^{({\rm ind}, D)}(g) \psi= \psi^\prime \nonumber \\
&&\psi^\prime (g^\prime)= \psi(g^{-1}g^\prime).
\end{eqnarray} 
It is clear that $G$ action preserves the covariance condition, 
and we have a 
UR of $G$ on ${\cal H}_{H}^{({\rm ind},D)}$. 

Whereas $D(\cdot)$ was assumed to be a UIR of $H$,  
${\cal D}_{H}^{({\rm ind}, D)}(\cdot)$ is in general reducible; so it is a direct 
sum of the various UIR's of $G$, each occurring with some multiplicity. These 
multiplicities are determined by the Reciprocity Theorem\cite{9} : Each UIR 
${\cal D}(\cdot)$ of $G$ appears in $ {\cal D}_{H}^{({\rm ind}, D)}(\cdot)$ 
as often as ${\cal D}(\cdot)$ contains $D(\cdot)$ upon restriction from $G$ to 
$H$. 

With this general background we now take up the specific case of $SU(3)$. 
The 
defining representation of this group is 
\begin{equation}
SU(3) = \{A= 3\times 3~ {\rm complex~ matrix} |A^\dagger A= I_{3\times 3}, 
{\rm det}~ A =1\},
\label{2.3} 
\end{equation}
with the group operation given by matrix multiplication. In this 
representation the eight hermitian generators are $\frac{1}{2} 
\lambda_{\alpha}, 
~~~\alpha=1,2,\cdots,8$, where the matrices $\lambda_{\alpha}$ and 
the structure constants 
$f_{\alpha\beta\gamma}$ occurring in the commutation relations
\begin{equation}
[\lambda_\alpha,\lambda_\beta]=2if_{\alpha\beta\gamma}\lambda_\gamma , 
~~~\alpha, \beta,\gamma=1,2,\cdots,8
\end{equation}
are all very well known\cite{10}.

A general UIR of $SU(3)$ is determined by two independent 
nonnegative integers $p$ and 
$q$, so it may be denoted as $(p,q)$. 
It is of dimension $d(p,q)= \frac{1}{2}
(p+1)(q+1)(p+q+2)$. The defining three dimensional UIR in $(\ref{2.3})$ is 
$(1,0)$; 
while the inequivalent complex conjugate UIR is $(0,1)$. In general the 
complex conjugate of $(p,q)$ is $(q,p)$; and the adjoint UIR is $(1,1)$ of 
dimension eight. Various choices of `magnetic quantum numbers' within a UIR 
may be made. The one corresponding to the canonical subgroup 
$SU(2)\times U(1)/Z_2=U(2)\subset SU(3)$ leads to the three quantum numbers 
$I, M, Y$ in standard notation. Here $I$ and $M$ are the isospin and 
magnetic quantum number  labels for a general UIR of $SU(2)$, while $Y$ is 
the eigenvalue of the (suitably normalised) $U(1)$ or hypercharge generator. 
The subgroups $SU(2)$ and $U(1)$ commute, and for definiteness we take 
$SU(2)$ to be the one acting on the first two dimensions of the three 
dimensions in the UIR $(1,0)$. The spectrum of `$I-Y$' multiplets 
present in 
the UIR $(p,q)$ can be described thus:  
\begin{equation}
I=\frac{1}{2}(r+s)~~~,~~~Y=r-s+\frac{2}{3}(q-p)~~,~~0~\leq~r\leq~p~,~
0~\leq~s~\leq~q .
\label{2.5}
\end{equation} 

\noindent
Thus for each  pair of integers $(r,s)$ in the above ranges, we have one 
$I-Y$ multiplet, with $M$ going over the usual $2I+1$ values 
$I,I-1,\cdots,-I+1,-I$. Then the orthonormal basis vectors for the UIR $(p,q)$ 
of $SU(3)$ may be written as $|p,q;I~ M~Y>$. This UIR can be realised via 
suitably constructed irreducible tensors. A tensor $T$ with $p$ indices 
belonging to the UIR $(1,0)$ and $q$ indices to the UIR $(0,1)$ 
is a collection of complex components $T_{k_1\cdots k_q}^{j_1\cdots j_p}$, 
$j$ and $k =1,2,3$, transforming under $A\in SU(3)$ by the rule 
\begin{equation}
{T^\prime}_{k_1\cdots k_q}^{j_1\cdots j_p}=
{A^{j_1}}_{l_1}\cdots {A^{j_p}}_{l_p}~{A^{k_1}}_{m_1}^*\cdots
{A^{k_q}}_{m_q}^*
~~T_{m_1\cdots m_q}^{l_1\cdots l_p}. 
\label{2.6}
\end{equation}
If in addition $T$ is completely symmetric separately in the superscripts 
and in the subscripts, and is traceless, i.e., contraction of any upper 
index with any lower index leads to zero, then all these properties are 
maintained under $SU(3)$ action and $T$ is an irreducible tensor. 
It then has 
precisely $d(p,q)$ independent components (in the complex sense); 
and the space of all such tensors carries the UIR $(p,q)$. The 
explicit transition from the tensor components 
$T_{k_1\cdots k_q}^{j_1\cdots j_p}$ to the canonical components 
$T_{IMY}^{(p,q)}$ may be found in \cite{11}

The regular representations of $SU(3)$ act on the space $L^2(SU(3))$, and in 
each of them the UIR $(p,q)$ appears $d(p,q)$ times. We 
shall not be concerned with this UR of $SU(3)$ in our work. Instead we give 
now the UIR contents of some selected induced UR's of $SU(3)$. For 
illustrative purposes we consider the following four subgroups
\begin{mathletters}
\begin{eqnarray}
U(1)\times U(1)&=&\{A={\rm diag}(e^{i(\theta_1+\theta_2)},
e^{i(\theta_1-\theta_2)},e^{-2i\theta_1})|0\leq\theta_1,\theta_2
\leq 2\pi\};
 \\
SU(2)&=&\{A=\pmatrix{a&{\bf 0}\cr
{\bf 0}&1\cr}| a\in SU(2)\};\\
U(2)&=&\{A=\pmatrix{u&{\bf 0}\cr
{\bf 0}&({\rm det}~u)^{-1}\cr}| u\in U(2)\};\\
SO(3)&=&\{A\in SU(3)|A^*=A\}.
\end{eqnarray} 
\end{mathletters}
In each case, we look at the induced UR of $SU(3)$ arising from the trivial 
one dimensional UIR of the subgroup. In the first two cases, in order to 
apply the Reciprocity Theorem, we can use the information in $(\ref{2.5})$ 
giving 
the $SU(2)\times U(1)/Z_2$ content of the UIR $(p,q)$ of $SU(3)$. Defining 
by a zero in the superscript the trivial UIR of the relevant subgroup, 
we have the results:
\begin{mathletters}
\begin{eqnarray}
{\cal D}_{U(1)\times U(1)}^{({\rm ind},0)}&=&\sum
\limits^{\infty}_{p,q=0,1,\cdots\atop{p=q~mod~ 3}}\oplus\;\;
 n_{p,q}~ (p,q)~~ ,~~ n_{p,q}={\rm min}(p+1,q+1);\\
{\cal D}_{SU(2)}^{({\rm ind},0)}&=&\sum\limits_{p,q=0,1,\cdots}
^{\infty}\oplus\;\;(p,q)~~ ;\label{2.8b}\\
{\cal D}_{U(2)}^{({\rm ind},0)}&=&\sum\limits_{p=0,1,\cdots}
^{\infty}\oplus\;\;(p,p)~~.
\end{eqnarray}
\end{mathletters}

The real dimensions of the corresponding coset spaces $SU(3)/U(1)\times U(1), 
SU(3)/SU(2)$ and $SU(3)/U(2)$ are $6$, $5$ and $4$ respectively. In the 
case of 
induction from the trivial UIR of $SO(3)$, we need to use the fact that the 
UIR $(p,q)$ of $SU(3)$ does not contain an $SO(3)$ invariant state if either 
$p$ or $q$ or both are odd, while it contains one such state if both $p$ and 
$q$ are even. Then we arrive at the reduction 
\begin{equation}
{\cal D}_{SO(3)}^{({\rm ind},0)}=\sum\limits_{r,s=0,1,\cdots}
^{\infty}\oplus\;\;(2r,2s)~~,
\end{equation}
with $SU(3)/SO(3)$ being of real dimension $5$.
 
From the above discussion we see that the induced UR 
${\cal D}_{SU(2)}^{({\rm ind},0)}$ of $SU(3)$ is particularly interesting in 
that it 
captures both the completeness and the economy properties of the Schwinger 
$SU(2)$ construction : each UIR of $SU(3)$ is present, exactly once. Thus we 
may call this a Generating Representation of $SU(3)$; it is much leaner than 
the regular representations.

\section{The minimal $SU(3)$ Schwinger oscillator construction}

An elementary oscillator operator construction of the $SU(3)$ generators is 
based on three independent pairs of annihilation and creation operators 
${\hat a}_j,{\hat a}_{j}^{\dagger}$ obeying
\begin{equation}
[{\hat a}_j,{\hat a}_{k}^{\dagger}]= \delta_{jk}~,~ [{\hat a}_j,{\hat a}_k]=
[{\hat a}_{j}^{\dagger},{\hat a}_{k}^{\dagger}]=0~,~ j,k=1,2,3~~.
\label{3.1}
\end{equation}
We write ${\cal H}^{(a)}$ for the Hilbert space on which these operators act 
irreducibly. The individual and total number operators are 
\begin{equation}
{\hat N}_{1}^{(a)}= {\hat a}_{1}^\dagger {\hat a}_1~~,~~ 
{\hat N}_{2}^{(a)}= {\hat a}_{2}^\dagger {\hat a}_2~~,~~
{\hat N}_{3}^{(a)}= {\hat a}_{3}^\dagger {\hat a}_3~~,~~
{\hat N}^{(a)}= {\hat a}_{j}^\dagger {\hat a}_j~~.
\end{equation}
If we now define the bilinear operators
\begin{equation}
Q_{\alpha}^{(a)}= \frac{1}{2}{\hat a}^\dagger \lambda_\alpha {\hat a}~~
,~~\alpha=1,2, 
\cdots,8,
\end{equation}
each $Q_{\alpha}^{(a)}$ is hermitian, and they obey the $SU(3)$ Lie algebra 
commutation relations
\begin{equation}
[Q_{\alpha}^{(a)},Q_{\beta}^{(a)}]=if_{\alpha\beta\gamma}~ Q_{\gamma}^{(a)}.
\end{equation}
In addition they conserve the total number operator:
\begin{equation}
[Q_{\alpha}^{(a)},{\hat N}^{(a)}]=0 
\end{equation}
Upon exponentiation of these generators we obtain a particular UR, 
${\cal U}^{(a)}(A)$ say, of $SU(3)$ acting on ${\cal H}^{(a)}$, under which 
the creation (annihilation) operators ${\hat a}_{j}^{\dagger}$ 
($ {\hat a}_j$) transform 
via the UIR $(1,0)$ ($(0,1)$):
\begin{eqnarray}
{\cal U}^{(a)}(A)~{\hat a}_{j}^{\dagger}~ {\cal U}^{(a)}(A)^{-1}&=&
 {A^{k}}_{j} {\hat a}_{k}^\dagger\nonumber\\
{\cal U}^{(a)}(A)~{\hat a}_j~{\cal U}^{(a)}(A)^{-1}&=&{A^{k}}_{j}^* {\hat a}_k
\end{eqnarray}
However upon reduction ${\cal U}^{(a)}(A)$ contains only the `triangular' 
UIR's $(p,0)$ of $SU(3)$, once each. In that sense this UR may be regarded 
as the `Generating Representation' for this subset of UIR's. For any given 
$p\geq 0$, the UIR $(p,0)$ is realised on that subspace ${\cal H}^{(p,0)}$ 
of ${\cal H}^{(a)}$ over which the total number operator ${\hat N}^{(a)}$ 
takes the eigenvalue $p$; and the connection between the tensor and the 
Fock space descriptions is given in this manner:
\begin{eqnarray}
\{T^{j_1\cdots j_p}\}\rightarrow|T>&=&T^{j_1\cdots j_p}~
{\hat a}_{j_1}^{\dagger} \cdots {\hat a}_{j_p}^{\dagger}|\underline{0}
>~ \in~ {\cal H}^{(p,0)}~ \subset ~{\cal H}^{(a)},
\nonumber\\
{\hat a}_j |\underline{0}>&=& 0;\nonumber\\
{\cal U}^{(a)}(A)~|T>&=& |T'>,\nonumber\\
{T^\prime}^{j_1\cdots j_p}&=& {A^{j_1}}_{l_1}\cdots {A^{j_p}}_{l_p} ~
T^{l_1\cdots l_p}.
\label{3.7}
\end{eqnarray} 
Therefore we have the (orthogonal) direct sum decompositions
\begin{eqnarray}
{\cal H}^{(a)}&=&\sum\limits_{p=0,1,\cdots}^{\infty}\oplus\;\;
{\cal H}^{(p,0)},\nonumber\\
{\cal H}^{(p,0)}&=&{\rm Sp}\{{\hat a}_{j_1}^{\dagger} \cdots 
{\hat a}_{j_p}^{\dagger}|\underline{0}>\},
\nonumber\\
{\cal U}^{(a)}&=&\sum\limits_{p=0,1,\cdots}^{\infty}
\oplus\;\;(p,0)
\label{3.8}
\end{eqnarray}

To be able to obtain the other UIR's as well, we bring in another 
independent 
triplet of oscillator operators ${\hat b}_j$ and 
${\hat b}_{j}^{\dagger}$ obeying the same 
commutation relations $(\ref{3.1})$ and commuting with ${\hat a}$'s and 
${\hat a}^\dagger$'s :
\begin{eqnarray}
[{\hat b}_j,{\hat b}_{k}^{\dagger}]&=& \delta_{jk}~,~[{\hat b}_j,{\hat b}_k] =
[{\hat b}_{j}^{\dagger},{\hat b}_{k}^{\dagger}]=0~,~ j,k=1,2,3~~,\nonumber\\ 
& & [{\hat a}_j~{\rm or}~ {\hat a}_{j}^{\dagger}~,
~ {\hat b}_k ~{\rm or}~ {\hat b}_{k}^{\dagger}] = 0
\end{eqnarray}
The corresponding Hilbert space is ${\cal H}^{(b)}$, and the $b$-type number 
operators are
\begin{equation}
{\hat N}_{1}^{(b)}= {\hat b}_{1}^\dagger {\hat b}_1~~,~~ 
{\hat N}_{2}^{(b)}= {\hat b}_{2}^\dagger {\hat b}_2~~,~~
{\hat N}_{3}^{(b)}= {\hat b}_{3}^\dagger {\hat b}_3~~,~~
{\hat N}^{(b)}= {\hat b}_{j}^\dagger {\hat b}_j~~.
\end{equation}
We define the $b$-type $SU(3)$ generators as 
\begin{equation}
Q_{\alpha}^{(b)}= -\frac{1}{2}{\hat b}^\dagger \lambda_{\alpha}^{*} 
{\hat b}~~,~~\alpha=1,2, 
\cdots,8,
\end{equation}
and they obey 
\begin{eqnarray}
[Q_{\alpha}^{(b)},Q_{\beta}^{(b)}]&=& if_{\alpha\beta\gamma}~ 
Q_{\gamma}^{(b)},\nonumber \\ {[Q_{\alpha}^{(b)},{\hat N}^{(b)}]} 
&=& 0 .
\end{eqnarray}
Exponentiation of these generators leads to a UR ${\cal U}^{(b)}(A)$ 
acting on ${\cal H}^{(b)}$, under which 
the creation (annihilation) operators ${\hat b}_{j}^{\dagger}$ 
($ {\hat b}_j$) transform 
via the UIR $(0,1)$ ($(1,0)$):
\begin{eqnarray}
{\cal U}^{(b)}(A)~{\hat b}_{j}^{\dagger}~ {\cal U}^{(b)}(A)^{-1}&=&
 {A^{k}}_{j}^*~ {\hat b}_{k}^\dagger\nonumber\\
{\cal U}^{(b)}(A)~{\hat b}_j~{\cal U}^{(b)}(A)^{-1}&=&{A^{k}}_{j}~ {\hat b}_k
\end{eqnarray}
Now this UR of $SU(3)$ contains each of the triangular UIR's $(0,q)$ for 
$q\geq 0$ once each, so it is a Generating Representation for this 
family of 
UIR's. For each $q\geq 0$, the UIR $(0,q)$ is realised on that subspace 
${\cal H}^{(0,q)}$ of ${\cal H}^{(b)}$ over which the total number operator 
${\hat N}^{(b)}$ takes the eigenvalue $q$. Analogous to $(\ref{3.7})$, the 
tensor-Fock space connection is now :
\begin{eqnarray}
\{T_{k_1\cdots k_q}\}\rightarrow|T>&=&T_{k_1\cdots k_q}~
{\hat b}_{k_1}^{\dagger} \cdots 
{\hat b}_{k_q}^{\dagger}|\underline{0}>~ \in~ {\cal H}^{(0,q)}~ 
\subset ~{\cal H}^{(b)},
\nonumber\\
{\hat b}_k |\underline{0}>&=& 0;\nonumber\\
{\cal U}^{(b)}(A)~|T>&=& |T'>,\nonumber\\
{T^\prime}_{k_1\cdots k_q}&=& {A^{k_1}}_{m_1}^*\cdots {A^{k_q}}_{m_q}^*
T_{m_1\cdots m_q}.
\label{3.14}
\end{eqnarray}
(The use of a common symbol $|\underline{0}>$ for the Fock ground states in 
${\cal H}^{(a)}$ and ${\cal H}^{(b)}$, and $|T>$ in $(\ref{3.7})$, 
$(\ref{3.14})$ should 
cause no confusion as the meanings are always clear from the context).
In place of $(\ref{3.8})$ we now have :
\begin{eqnarray}
{\cal H}^{(b)}&=&\sum\limits_{q=0,1,\cdots}^{\infty}\oplus\;\;{\cal H}^{(0,q)},\nonumber\\
{\cal H}^{(0,q)}&=&{\rm Sp}\{{\hat b}_{k_1}^{\dagger} \cdots 
{\hat b}_{k_q}^{\dagger}|\underline{0}>\},
\nonumber\\
{\cal U}^{(b)}&=&\sum\limits_{q=0,1,\cdots}^{\infty}
\oplus\;\;(0,q)
\end{eqnarray}

From these considerations it is clear that if we want to obtain all the UIR's 
$(p,q)$ of $SU(3)$, missing none, the minimal scheme is to use all six 
independent oscillators ${\hat a}_j, {\hat a}_{j}^\dagger, 
{\hat b}_j, {\hat b}_{j}^{\dagger}$ and define the $SU(3)$ generators\cite{12} 
\begin{equation}
Q_\alpha= Q_{\alpha}^{(a)} +Q_{\alpha}^{(b)}.
\end{equation} 
They act on the product Hilbert space ${\cal H}= {\cal H}^{(a)}
\times {\cal H}^{(b)}$, 
of course obey the $SU(3)$ commutation relations, and upon exponentiation 
lead to the UR ${\cal U}(A)= {\cal U}^{(a)}(A) \times {\cal U}^{(b)} (A)$. 
However, as we see in a moment, while each UIR $(p,q)$ is certainly present 
in ${\cal U}(A)$, it occurs infinitely many times. A systematic group 
theoretic procedure to handle this multiplicity, based on the non compact    
group $Sp(2,R)$, will be set up below. The tensor-Fock space connection 
is now 
given as follows. To an irreducible tensor 
$T_{k_1\cdots k_q}^{j_1\cdots j_p}$ which is symmetric and traceless 
and so `belongs' to the UIR $(p,q)$ we associate the vector $|T> \in {\cal H}$ 
by 
\begin{eqnarray}
|T> &=&{T}_{k_1\cdots k_q}^{j_1\cdots j_p}~
{\hat a}_{j_1}^{\dagger}\cdots {\hat a}_{j_p}^{\dagger}
{\hat b}_{k_1}^{\dagger}\cdots 
{\hat b}_{k_q}^{\dagger}|\underline{0},\underline{0}>~\in 
{\cal H}^{(p,0)}\times {\cal H}^{(0,q)}
~\subset~{\cal H}, \nonumber\\
{\hat a}_j|\underline{0},\underline{0}>&=& 
{\hat b}_j |\underline{0},\underline{0}>=0,\nonumber\\
{\cal U}(A)|T>&=&|T^\prime>, 
\label{3.17}
\end{eqnarray}
the components of $T^\prime$ being given by $(\ref{2.6})$. While this 
vector $|T>$ is 
certainly a simultaneous  eigenvector of the two number operators 
${\hat N}^{(a)}, {\hat N}^{(b)}$ with eigenvalues $p,q$ respectively, the 
tracelessness of the tensor ${T}_{k_1\cdots k_q}^{j_1\cdots j_p}$ 
implies that 
(unless at least one of $p$ and $q$ vanishes) we do not get all such 
independent vectors in ${\cal H}$. This aspect is further clarified 
below. On 
the other hand if we drop the tracelessness condition and retain only 
symmetry, we do span all of ${\cal H}^{(p,0)} \times {\cal H}^{(0,q)}$ via 
$(\ref{3.17})$. 

The decomposition of ${\cal U}(A)$ into UIR's, and the counting of 
multiplicities, is accomplished by appealing to the Clebsch-Gordan 
Series for 
the product of two triangular UIR's $(p,0)$ and $(0,q)$\cite{13}:
\begin{equation}
(p,0) \times (0,q) = (p,q) \oplus (p-1,q-1)\oplus
(p-2,q-2)\oplus\ldots\oplus (p-r,q-r)~,~ r={\rm min}(p,q)
\label{3.18}
\end{equation}
Therefore at the Hilbert space level one has the orthogonal subspace 
decomposition
\begin{eqnarray}
{\cal H} &=& {\cal H}^{(a)}~\times~{\cal H}^{(b)}\nonumber\\
&=&\left( \sum\limits_{p=0,1,\cdots}^{\infty}\oplus\;\;
{\cal H}^{(p,0)}\right)
~\times~\left( \sum\limits_{q=0,1,\cdots}^{\infty}\oplus\;\;
{\cal H}^{(0,q)}\right)\nonumber\\
&=& \sum\limits_{p,q=0,1,\cdots}^{\infty}\oplus\;\;
{\cal H}^{(p,0)}~\times~{\cal H}^{(0,q)},\nonumber\\
{\cal H}^{(p,0)}~\times~{\cal H}^{(0,q)}&=& 
\sum\limits_{\rho=0,1,\cdots}^{r}\oplus\;\;
{\cal H}^{(p-\rho,q-\rho~;~\rho)}~,~r={\rm min}(p,q).
\label{3.19}
\end{eqnarray} 
Here ${\cal H}^{(p-\rho,q-\rho~;~\rho)}$ is that unique subspace of  
${\cal H}^{(p,0)}~\times~{\cal H}^{(0,q)}$ carrying the UIR $(p-\rho,q-\rho)$ 
present on the right hand side of $(\ref{3.18})$. All vectors in 
${\cal H}^{(p-\rho,q-\rho~;~\rho)}$ are eigen vectors of ${\hat N}^{(a)}$ 
and ${\hat N}^{(b)}$ with eigenvalues $p$ and $q$ respectively; and if the 
tensor $T$ in $(\ref{3.17})$ is assumed traceless, only vectors in 
${\cal H}^{(p,q~;~0)}~\subset~{\cal H}^{(p,0)}~\times~{\cal H}^{(0,q)} $ are 
obtained on the right in that equation.

Focussing on a given UIR $(p,q)$, we see that it appears once each in 
${\cal H}^{(p,0)}~\times~{\cal H}^{(0,q)}, 
{\cal H}^{(p+1,0)}~\times~{\cal H}^{(0,q+1)}, \cdots$ in the respective 
irreducible subspaces ${\cal H}^{(p,q~;~0)}, {\cal H}^{(p,q~;~1)}, \cdots $. 
Thus it is the leading piece in ${\cal H}^{(p,0)}~\times~
{\cal H}^{(0,q)}$, the next to the leading piece in 
${\cal H}^{(p+1,0)}~\times~{\cal H}^{(0,q+1)}$, and so on. Therefore the 
decomposition $(\ref{3.19})$ of ${\cal H}$ can be presented in the alternative 
manner
\begin{equation}
{\cal H}=\sum\limits_{p,q=0,1.\cdots}^{\infty}\oplus
\sum\limits_{\rho=0,1.\cdots}^{\infty}\oplus\;\;
{\cal H}^{(p,q~;~\rho)},\; {\cal H}^{(p,q~;~\rho)}~\subset~
{\cal H}^{(p+\rho,0)}~\times~{\cal H}^{(0,q+\rho)}
\end{equation}
each ${\cal H}^{(p,q~;~\rho)}$ carrying the same UIR $(p,q)$. Thus the index 
$\rho$ is an (orthogonal) multiplicity label with an infinite number of 
values. For $\rho \neq \rho^\prime$, ${\cal H}^{(p,q~;~\rho^\prime)}$ and 
${\cal H}^{(p,q~;~\rho)}$ are mutually orthogonal. This is also  evident 
as ${\hat N}^{(a)}=p+\rho^\prime$, ${\hat N}^{(b)}=q+\rho^\prime$ in the 
former and ${\hat N}^{(a)}=p+\rho$, ${\hat N}^{(b)}=q+\rho$ in the latter. 

We now introduce the group $Sp(2,R)$ to handle in a systematic way the 
multiplicity index $\rho$. The hermitian generators of $Sp(2,R)$ and their 
commutation relations are\cite{14}
\begin{eqnarray}
J_0 &=& \frac{1}{2}({\hat N}^{(a)}+{\hat N}^{(b)} +3),\nonumber\\
K_1 &=& \frac{1}{2}({\hat a}_{j}^{\dagger} {\hat b}_{j}^{\dagger} 
+ {\hat a}_j {\hat b}_j), \nonumber\\
K_2 &=& -\frac{i}{2}({\hat a}_{j}^{\dagger} {\hat b}_{j}^{\dagger} 
- {\hat a}_j {\hat b}_j); \nonumber\\
{[J_0, K_1]}&=& {iK_2, [J_0,K_2]=-iK_1, [K_1,K_2]= -iJ_0.}
\label{3.21}
\end{eqnarray}
Using the raising and lowering combinations $K_{\pm}=K_1 \pm iK_2$ we have :
\begin{eqnarray}
K_{+} &=& {\hat a}_{j}^{\dagger} {\hat b}_{j}^{\dagger}, 
K_{-}=K^{\dag}_{+}={\hat a}_j {\hat b}_j;\nonumber\\
\protect[J_0,K_{\pm}\protect]&=& \pm K_{\pm},
 \protect[K_+,K_-\protect]= -2J_0.
\end{eqnarray}

The significance of this construction is that the two groups $SU(3)$ and 
$Sp(2,R)$, both acting unitarily on ${\cal H}$, commute with one another :
\begin{equation}
[ J_0~{\rm or}~K_1~{\rm or}~K_2, Q_\alpha]=0
\label{3.23}
\end{equation}
It is this that helps us handle the multiplicity of occurrences of each 
$SU(3)$ UIR $(p,q)$ in ${\cal H}:\rho$ becomes a `magnetic quantum 
number' within a suitable UIR of $Sp(2,R)$.

The family of (infinite dimensional) UIR's of $Sp(2,R)$ relevant here is the 
positive discrete family $D_{k}^{(+)}$, labelled by $k=1/2,1,3/2,2\cdots$ 
(Actually we encounter only $k \geq 3/2$). Within the UIR $D_{k}^{(+)}$ we 
have an orthonormal basis $|k,m>$ on which the generators act as 
follows\cite{15}:
\begin{eqnarray}
J_0 |k,m> &=& m|k,m>~~,~~m=k,k+1,k+2,\cdots\nonumber\\
K_{\pm} |k,m> &=& \sqrt{(m\pm k)(m \mp k \pm 1)}|k,m\pm 1>
\end{eqnarray}
From these follow the useful results
\begin{mathletters}
\begin{eqnarray}
K_{1}^{2}+ K_{2}^{2}-J_{0}^{2}&=&k(1-k),\\
|k,m> &=& \sqrt{\frac{(2k-1)!}{(m-k)!(m+k-1)!}} K_{+}^{m-k}|k,k>,\label{3.25b}
\\
 K_{+}^{m-k} K_{-}^{m-k}|k,m> &=& \frac{(m-k)!(m+k-1)!}{(2k-1)!}|k,m>.
\label{3.25c}
\end{eqnarray}
\end{mathletters}
Going back to the generators $(\ref{3.21})$ it is clear that on all of 
${\cal H}^{(p,0)}~\times~{\cal H}^{(0,q)}$, and so on each 
${\cal H}^{(p-\rho,q-\rho~;~\rho)}$, $J_0$ has the eigenvalue 
$\frac{1}{2}(p+q+3)$; therefore on ${\cal H}^{(p,q~;~\rho)}$ it has the 
eigenvalue $\frac{1}{2}(p+q+3)+\rho$. It is also clear that action by 
$K_{\pm}$ on ${\cal H}^{(p,0)}~\times~{\cal H}^{(0,q)}$ leads to a 
subspace of ${\cal H}^{(p\pm 1,0)}~\times~{\cal H}^{(0,q\pm 1)}$. 
Therefore because of $(\ref{3.23})$ we see that $K_{\pm}$ acting on   
${\cal H}^{(p,q~;~\rho)}$ yield ${\cal H}^{(p,q~;~\rho \pm 1)}$. 
Of course ${\cal H}^{(p,q~;~0)}$ is annihilated by $K_{-}$. 

Reflecting all this we see that an orthonormal basis for ${\cal H}$ can 
be set up labelled as follows:
\begin{eqnarray}
|p,q; IMY; m> &:& p,q=0, 1,2,\ldots;\nonumber\\
m&=&k, k+1, k+2,\ldots,\nonumber\\
k&=& \frac{1}{2}(p+q+3) ;\nonumber\\
N^{(a)} = p+m-k&,& N^{(b)}=q+m-k .
\label{3.26}
\end{eqnarray}

\noindent
Since $k$ is determined in terms of $p$ and $q$, we do not include 
it as an additional label in the basis kets above.
(The ranges for $I,M,Y$ within the $SU(3)$ UIR $(p,q)$ are given in 
$(\ref{2.5})$) The $SU(3)$  UIR labels $p,q$ determine $k$ and so the 
associated UIR $D_{k}^{(+)}$ of $Sp(2,R)$. For fixed $p,q$ 
 as $I,M,Y,m$ vary we get a set of states carrying 
the UIR $(p,q)~\times D_{k}^{(+)}$ of $SU(3)~\times Sp(2,R)$. 
We can now appreciate the following relationships :
\begin{mathletters}
\begin{eqnarray}
{{\cal H}^{(p,q~;~\rho)}} &=& {{\rm Sp}\{|p,q;I M Y;
k+\rho>|I M Y {\rm varying}\},}\nonumber\\
&&{ \rho=0,1,2,\cdots};\label{3.27b}\\
{{\cal H}^{(p,q~;~\rho)}} &=&{K_{+}^{\rho}{\cal H}^{(p,q~;~0)};}\\
 K_{-}{\cal H}^{(p,q~;~0)} &=&{0.}
\end{eqnarray}
\end{mathletters} 

Therefore the null space of $K_{-}$ within ${\cal H}$ is the subspace
\begin{eqnarray}
{\cal H}_0 &=& \sum\limits_{p,q=0,1,\cdots}^{\infty}\;\;
\oplus {\cal H}^{(p,q;0)},\nonumber\\
&&\nonumber\\
&=& Sp\{|p,q; IMY; k>| p,q,IMY\;\mbox{varying}\},
\label{3.28}
\end{eqnarray}
and we see that  the UR ${\cal U}(A)$ of $SU(3)$ on ${\cal H}$ when 
restricted to ${\cal H}_0$ gives a UR ${\cal D}_0$ which is multiplicity 
free and includes every UIR of 
$SU(3)$. It is thus identical in structure to the induced representation 
${\cal D}_{SU(2)}^{({\rm ind},0)}$ in $(\ref{2.8b})$. We see how the use of 
$Sp(2,R)$ helps us isolate ${\cal H}_0$ in a neat manner.                     

In addition to the subspaces ${\cal H}^{(p,q;\rho)}, {\cal H}_0$
of ${\cal H}$ defined above, it is also useful to define the series of
mutually orthogonal infinite dimensional subspaces
     \begin{eqnarray}
     {\cal H}^{(p,q)} &=& \sum\limits^{\infty}_{\rho=0}\oplus
     {\cal H}^{(p,q;\rho)}\nonumber\\
     &=& Sp\{|p,q; IMY; m>|IMYm\;\mbox{varying}\},\nonumber\\
     &&p,q=0, 1, 2, \ldots .
     \end{eqnarray}

\noindent
Thus the infinity of occurrences of the $SU(3)$ UIR $(p,q)$
are collected together in ${\cal H}^{(p,q)}$.

In the appendix we give explicit formulae for the state vectors
$|p,q;IMY;m>$ as functions of the operators ${\hat a}^{\dag}_j, {\hat b}
^{\dag}_j$ acting on the Fock vacuum $|\underline{0},\underline{0}>$.

\section{The Bargmann Representation}

For some purposes the use of the Bargmann representation of the canonical 
commutation relations is more convenient than the Fock space 
description\cite{16}. 
We outline the definitions of ${\cal H}$ and the $SU(3)$ UR 
${\cal U}(A)={\cal U}^{(a)}(A)~\times~{\cal U}^{(b)}(A)$ in this language, 
and then turn to the problem of isolating the subspace ${\cal H}_0$ 
in ${\cal H}$. 

Vectors in ${\cal H}$ correspond to entire functions $f({\underline z}, 
{\underline w})$ in six independent complex variables ${\underline z}= (z_j), 
{\underline w} =(w_j), j=1,2,3$ with the squared norm defined as 
\begin{equation}
||f||^2 = \int \prod_{j=1}^{3}\left(\frac{d^2 z_j}{\pi}\right)
\left(\frac{d^2 w_j}{\pi}\right) e^{-z^\dagger z-w^\dagger w}
 |f({\underline z}, {\underline w})|^2 
\label{4.1}
\end{equation}
Any such $f({\underline z}, {\underline w})$ has a unique Taylor series 
expansion
\begin{equation}
f({\underline z}, {\underline w})= \sum_{p,q=0,1,\cdots}^{\infty}
f_{k_1\cdots k_q}^{j_1\cdots j_p} z_{j_1}\cdots z_{j_p}~w_{k_1}\cdots 
w_{k_q}, 
\label{4.2}
\end{equation}
involving the tensor components $ f_{k_1\cdots k_q}^{j_1\cdots j_p}$ 
separately symmetric in the 
superscripts and the subscripts. In terms of these the squared norm is 
\begin{equation}
||f||^2= \sum\limits_{p,q=0,1,\cdots}^{\infty}~p!~q!~ 
{f_{k_1\cdots k_q}^{j_1\cdots j_p}}^* 
~f_{k_1\cdots k_q}^{j_1\cdots j_p}
\label{4.3}
\end{equation}
The operators ${\hat a}_j,{\hat a}_{j}^\dagger, 
{\hat b}_j,{\hat b}_{j}^\dagger$ act on 
$f({\underline z}, {\underline w})$ as follows:
\begin{equation}
{\hat a}_j\rightarrow \frac{\partial}{\partial z_j}, 
{\hat a}_{j}^\dagger \rightarrow z_j, 
{\hat b}_j\rightarrow \frac{\partial}{\partial w_j}, 
{\hat b}_{j}^\dagger \rightarrow w_j.
\end{equation}
The UR ${\cal U}(A)$ of $SU(3)$ acts very simply via point transformations:
\begin{equation}
({\cal U}(A)f)({\underline z}, {\underline w}) = 
f(A^{-1}{\underline z}, A^{-1*}{\underline w}).
\label{4.5}
\end{equation}
The $Sp(2,R)$ generators are particularly simple:
\begin{eqnarray}
{J_0} &=&{\frac{1}{2}\left(z_j\frac{\partial}{\partial z_j}
+w_j\frac{\partial}{\partial w_j}+ 3\right),}\nonumber\\
{K_{+}}&=& {z_j w_j \equiv {\underline z}\cdot {\underline w},}\nonumber\\
{K_{-}}&=&{ \frac{\partial^2}{\partial z_j \partial w_j}\equiv 
\frac{\partial}{\partial {\underline z}}\cdot\frac{\partial}{\partial
{\underline w}}.}
\end{eqnarray}
We will use these below. 

It is clear that the terms in $(\ref{4.2}), (\ref{4.3})$ for fixed $p$ and $q$
are contributions from ${\cal H}^{(p,0)}~\times{\cal H}^{(0,q)}$. The action 
by $K_{+}$ obeys:
\begin{equation}
f({\underline z}, {\underline w})\in {\cal H}^{(p,0)}~\times{\cal H}^{(0,q)} 
\rightarrow K_{+}f({\underline z}, {\underline w})=
{\underline z}\cdot {\underline w}f({\underline z}, {\underline w}) 
\in {\cal H}^{(p+1,0)}~\times{\cal H}^{(0,q+1)} 
\end{equation} 
On the other hand action by $K_{-}$ is the analytic equivalent of taking the 
trace: starting with (4.2) we get
\begin{equation}
K_{-}f({\underline z}, {\underline w})
= \sum_{p,q=0,1,\cdots}^{\infty}~pq~
f_{jk_1\cdots k_{q-1}}^{jj_1\cdots j_{p-1}}~ z_{j_1}\cdots z_{j_{p-1}}
~w_{k_1}\cdots w_{k_{q-1}}.
\end{equation}
From these and earlier remarks we can see that the correspondences between 
(symmetric, traceless) tensors, entire functions, and subspaces of 
${\cal H}$ are:
\begin{mathletters}
\begin{eqnarray}
{{\cal H}^{(p,0)}~\times{\cal H}^{(0,q)}}&&{\leftrightarrow 
\left\{ f_{k_1\cdots k_q}^{j_1\cdots j_p}\right\}
\leftrightarrow
f({\underline z}, {\underline w}):}\nonumber\\ 
&&{f(\lambda{\underline z}, \mu{\underline w})=\lambda^p \mu^q 
f({\underline z}, {\underline w});}\\
{f({\underline z}, {\underline w}) \in {\cal H}^{(p,q~;~\rho)}
\Leftrightarrow f({\underline z}, {\underline w})}&&=
(\underline {z}\cdot\underline {w})^{\rho}~
f_0({\underline z}, {\underline w}),
f_0({\underline z}, {\underline w}) \in
 {\cal H}^{(p,q~;~0)} \subset 
 {\cal H}^{(p,0)}~\times{\cal H}^{(0,q)},\nonumber\\
&&{\frac{\partial}{\partial {\underline z}}\cdot\frac{\partial}{\partial
{\underline w}}f_0({\underline z}, {\underline w})=0}
\label{4.9b}
\end{eqnarray}
\end{mathletters}
Thus traceless symmetric tensors of type $(p,q)$ are in correspondence with 
entire functions $f_0({\underline z}, {\underline w})$ of degrees of 
homogeneity $p$ and $q$ respectively, obeying the partial differential 
equation $(\ref{4.9b})$. Alternatively, given any 
$f({\underline z}, {\underline w})\in {\cal H}^{(p,0)}~\times{\cal H}^{(0,q)}$, there is a unique `traceless' part $f_0({\underline z}, {\underline w})$ 
belonging to the leading subspace ${\cal H}^{(p,q~;0)}$ and annihilated by 
$K_{-}$. Thus `trace removal' can be accomplished by analytical means. 
We now 
give the procedure to pass from  $f({\underline z}, {\underline w})$ to   
 $f_0({\underline z}, {\underline w})$.

For any $f({\underline z}, {\underline w}) \in 
{\cal H}^{(p,0)}~\times{\cal H}^{(0,q)}$ we can easily establish the general 
formula
\begin{eqnarray}
K_-\{(\underline{z}\cdot \underline{w})^n~ K_{-}^{n}~
f(\underline{z}, \underline{w})\} &=& n(p+q+2-n) 
(\underline {z}\cdot \underline {w})^{n-1}~ K_{-}^{n}
f(\underline {z}, \underline {w})\nonumber\\
&+& (\underline{z}\cdot\underline{w})^n\;K_{-}^{n+1}\; f(\underline{z},
\underline{w})
\label{4.10}
\end{eqnarray}
We try for $f_0({\underline z}, {\underline w})$ the expression
\begin{equation}
f_0({\underline z}, {\underline w})= 
f({\underline z}, {\underline w})-\sum_{n=1,2,\cdots} \alpha_n 
({\underline z}\cdot {\underline w})^n K_{-}^{n}
f({\underline z}, {\underline w})
\end{equation}
and get using $(\ref{4.10})$ (and omitting the arguments
${\underline z}, {\underline w}$):
\begin{eqnarray}
{K_{-}f_0 = K_{-}f}&& - {(p+q+1)\alpha_1 K_{-}f}\nonumber\\
&&{ - 
\sum_{n=1,2,\cdots}\{\alpha_n+(n+1)(p+q+1-n)\alpha_{n+1}\}
({\underline z}\cdot {\underline w})^n K_{-}^{n+1}f}.
\end{eqnarray}
We can therefore attain $K_{-}f_0 =0$ by choosing
\begin{equation}
\alpha_n = (-1)^{n-1}\frac{(p+q+1-n)!}{n!(p+q+1)!},n=1,2,\ldots.
\end{equation}
Therefore for any (bihomogeneous) polynomial $f({\underline z}, 
{\underline w})
\in {\cal H}^{(p,0)}~\times{\cal H}^{(0,q)}$ the leading traceless part 
annihilated by $K_{-}$ is an element $f_0(\underline{z},\underline{w})$
 in ${\cal H}^{(p,q~;~0)}$ :
\begin{equation}
f_0({\underline z}, {\underline w})= f({\underline z}, {\underline w})
- \sum_{n=1,2,\cdots}(-1)^{n-1}\frac{(p+q+1-n)!}{n!(p+q+1)!}
(\underline{z}\cdot\underline{w})^n\;
K_{-}^{n} f({\underline z}, {\underline w})
\label{4.14}
\end{equation} 

This result can be extended and expressed in the Fock space language.
Any $|\psi> \in {\cal H}^{(p,0)}~\times{\cal H}^{(0,q)}$ has a unique 
orthogonal decomposition into various parts belonging to various UIR's 
of $SU(3)$; using $(\ref{3.25c})$ this reads: 
\begin{eqnarray}
{|\psi>}&{ \in}&{ {\cal H}^{(p,0)}~\times{\cal H}^{(0,q)}
= {\cal H}^{(p,q~;0)}\bigoplus {\cal H}^{(p-1,q-1~;~1)}\bigoplus 
{\cal H}^{(p-2,q-2~;~2)}\bigoplus\cdots :}\nonumber\\
{|\psi>}&{=}&{|\psi_0>+|\psi_1>+|\psi_2>+\cdots ,}\nonumber\\
{|\psi_0>}&{\in}&{{\cal H}^{(p,q~;0)}, K_{-}|\psi_0> =0;}\nonumber\\
{|\psi_1>}&{=}&{K_{+}|\phi_1> \in {\cal H}^{(p-1,q-1~;1)},}\nonumber\\
{|\phi_1>}&{=}&{ \frac{(p+q)!}{1!(p+q+1)!} 
~K_{-}|\psi_1> \in~ {\cal H}^{(p-1,q-1~;0),}}\nonumber\\
{K_{-}^{2}|\psi_1>}&{=}& {0;}\nonumber\\
{|\psi_2>}&{=}&{K_{+}^{2}|\phi_2> \in {\cal H}^{(p-2,q-2~;2)},}\nonumber\\
{|\phi_2>}&{=}&{ \frac{(p+q-2)!}{2!(p+q)!} 
~K_{-}^2|\psi_2> \in~ {\cal H}^{(p-2,q-2~;0),}}\nonumber\\
{K_{-}^{3}|\psi_2>}&{=}& {0; \cdots }
\end{eqnarray}
The `leading' piece in $|\psi>$ is thus 
\begin{eqnarray}
{|\psi_0>}&{=}&{|\psi>-|\psi_1>-|\psi_2>-\cdots}\nonumber\\
          &{=}&{|\psi>-{\underline {\hat a}}^\dagger \cdot
{\underline {\hat b}}^\dagger
|\phi>,}\nonumber\\
{|\phi>}&{=}&{|\phi_1>+{\underline {\hat a}}^\dagger 
\cdot{\underline {\hat b}}^\dagger
|\phi_2> +\cdots \in {\cal H}^{(p-1,0)}~\times{\cal H}^{(0,q-1)}.}   
\end{eqnarray} 
We can now infer that if to begin with we had 
$|\psi>={\underline {\hat a}}^\dagger \cdot
{\underline {\hat b}}^\dagger |\phi>$ for some 
$\phi\in {\cal H}^{(p-1,0)}~\times{\cal H}^{(0,q-1)}$ then $|\psi_0>$ 
necessarily vanishes:
\begin{equation}
|\psi>={\underline {\hat a}}^\dagger \cdot
{\underline {\hat b}}^\dagger |\phi> \Leftrightarrow |\psi_0>=0
\end{equation}
In the Bargmann description this means in terms of $(\ref{4.14})$
\begin{equation}
f({\underline z}, {\underline w})={\underline z}\cdot {\underline w}
g({\underline z}, {\underline w}) \Leftrightarrow 
f_0({\underline z}, {\underline w})=0,
\end{equation}
a result which can be directly verified with some effort. 

The subspace ${\cal H}_0 \subset {\cal H}$ identified in $(\ref{3.28})$ is 
describable in the Bargmann language as follows:
\begin{equation}
{\cal H}_0 =\{f({\underline z}, {\underline w})\;\in\;{\cal H}|
\frac{\partial}{\partial {\underline z}}\cdot\frac{\partial}{\partial
{\underline w}}f({\underline z}, {\underline w})=0\}
\label{4.19}
\end{equation}
In the Taylor series expansion $(\ref{4.2})$ for such 
$f({\underline z}, {\underline w})$, the tensors 
$f_{k_1\cdots k_q}^{j_1\cdots j_p}$ are traceless and vice versa. The 
squared norm and $SU(3)$ action are given for ${\cal H}_0$ by $(\ref{4.3})$ 
and $(\ref{4.5})$ respectively.

\section{The UR ${\cal D}_{SU(2)}^{({\rm ind},0)}$ of $SU(3)$}

The Hilbert space ${\cal H}_{SU(2)}^{({\rm ind},0)}$ carrying the UR 
${\cal D}_{SU(2)}^{({\rm ind},0)}$ of $SU(3)$ consists of single component (scalar) 
complex functions on the coset space $SU(3)/SU(2)$. This coset space is the 
unit sphere in three dimensional complex space $C^3$, with the natural norm 
and $SU(3)$ action. Temporarily omitting the superscript zero and subscript 
$SU(2)$ for simplicity, we have:
\begin{eqnarray}
{{\cal H}^{({\rm ind})}}&=&{\{\psi({\underline \xi})\in C, {\underline \xi} \in C^3|
||\psi||^2 = \int \prod_{j=1}^{3}\left(\frac{d^2 \xi_j}{\pi}\right)
\delta(\xi^\dagger \xi-1) |\psi({\underline \xi})|^2\},}\nonumber\\
{({\cal D}^{({\rm ind})}(A)\psi)({\underline \xi})}&=&{\psi(A^{-1}{\underline \xi})}
\label{5.1}
\end{eqnarray}
Clearly only the values of $\psi({\underline \xi})$ for $\xi^\dagger \xi= 1$ 
are relevant. For a general $\psi({\underline \xi})$ with a Taylor series 
expansion we write
\begin{equation}
\psi({\underline \xi})= \sum_{p,q=0,1,\cdots}^{\infty}
\psi_{k_1\cdots k_q}^{j_1\cdots j_p} \xi_{j_1}\cdots \xi_{j_p}~
\xi_{k_1}^{*}\cdots \xi_{k_q}^{*}, 
\label{5.2}
\end{equation}
(Strictly speaking, such an expansion holds only for 
$\psi({\underline \xi})$ in some dense subset of ${\cal H}^{({\rm ind})}$). 
We note 
that here $\psi({\underline \xi})$ is not an entire function of $\xi_j$, 
and since $\xi^\dagger \xi= 1$, the tensor components 
$\psi_{k_1\cdots k_q}^{j_1\cdots j_p}$ may be assumed to be traceless apart 
from being symmetric. Then they determine $\psi({\underline \xi})$ uniquely 
and vice versa. 

To express the inner product $(\phi,\psi)$ for general $\phi, \psi 
\in {\cal H}^{({\rm ind})}$ in terms of their tensor components, we need to 
evaluate 
\begin{equation}
I_{j\cdots m \cdots}^{k\cdots l \cdots} = 
\int \prod_{j=1}^{3}\left(\frac{d^2 \xi_j}{\pi}\right)
\delta(\xi^\dagger \xi-1)~
\xi_{j_1}\cdots \xi_{j_p}~(\xi_{k_1}\cdots \xi_{k_q})^* 
~(\xi_{l_1}\cdots \xi_{l_{p^{\prime}}})^*~(\xi_{m_1}\cdots 
\xi_{m_{q^{\prime}}}),
\end{equation}
for general $p, q, p^\prime, q^\prime$ and indices $j, k, l, m$. Using 
$SU(3)$ invariance and symmetry, we see that the result must be expressible 
in terms of products of Kronecker deltas. Combining this with the 
tracelessness of the tensor components of $\phi$ and $\psi$, we can check 
first that we need only consider the case $p=p^\prime, q=q^\prime$; and next 
that 
\begin{equation}
I_{j\cdots m \cdots}^{k\cdots l \cdots} = {\cal N}
\sum_{P\in S_p}\sum_{Q\in S_q} 
\delta_{j_1}^{l_{P(1)}}\cdots \delta_{j_p}^{l_{P(p)}} 
~ \delta_{m_{Q(1)}}^{k_1}\cdots  \delta_{m_{Q(q)}}^{k_q}+\cdots
\end{equation}  
Here ${\cal N}$ is a normalising factor, and the dots denote terms with 
factors $\delta_{j}^{k}$ or $\delta_{l}^{m}$ or both. Again the latter 
can be ignored. The factor ${\cal N}$ can be computed say by setting 
all $j=l=1$ and all $k=m=2$ :
\begin{equation}
{\cal N} = \frac{1}{(p+q+2)!}.
\end{equation}
We then get the result for any $\phi,\psi \in {\cal H}^{({\rm ind})}$:
\begin{equation}
(\phi,\psi)=\sum_{p,q=0,1,\cdots} \frac{p!q!}{(p+q+2)!}  
{\phi_{k_1\cdots k_q}^{j_1\cdots j_p}}^* 
{\psi_{k_1\cdots k_q}^{j_1\cdots j_p}}
\label{5.6}
\end{equation}

With these results, all  details of the induced UR 
${\cal D}_{SU(2)}^{({\rm ind},0)}$ of $SU(3)$ are in hand : the Hilbert space 
${\cal H}_{SU(2)}^{({\rm ind},0)}$ in $(\ref{5.1})$, the expression $(\ref{5.6})$ 
for inner products, and the $SU(3)$ action as in $(\ref{5.1})$.

\section{Equivalence map}

The full equivalence of the two UR's of $SU(3)$, one on the subspace 
${\cal H}_0 
\subset {\cal H}$ based on the six oscillator Schwinger construction of 
Section III, and the other the induced representation 
${\cal D}_{SU(2)}^{({\rm ind},0)}$, can now be set up. The tensor component 
expressions $(\ref{4.2}), (\ref{5.2})$ for vectors, and $(\ref{4.3}), 
(\ref{5.6})$ for inner 
products, determine the one-to-one map to achieve this in full detail :
\begin{eqnarray}
{f({\underline z}, {\underline w})=
\{f_{k_1\cdots k_q}^{j_1\cdots j_p}\}}&&{ \in {\cal H}_0
\leftrightarrow \psi({\underline \xi})
=\{\psi_{k_1 \cdots k_q}^{j_1\cdots j_p}\} \in {\cal H}^{({\rm ind})}}:
\label{6.1}
\nonumber\\
{\psi_{k_1\cdots k_q}^{j_1\cdots j_p}}&&={\sqrt{(p+q+2)!}
f_{k_1\cdots k_q}^{j_1\cdots j_p}}, p,q=0,1,\ldots.
\end{eqnarray}
The two inner products then match, and the $SU(3)$ actions given in 
$(\ref{4.5}), (\ref{5.1})$ on $f({\underline z}, {\underline w})$ and  
$\psi({\underline \xi})$ also match. 

It is worth emphasising here the two different arguments leading to the 
tracelessness of the symmetric tensors on the two sides of $(\ref{6.1})$. 
In the 
case of the left hand side, the reason is that the argument of 
$\psi({\underline \xi})$ obeys the constraint $\xi^\dagger \xi =1$. As for 
the right hand side, it happens because entire functions
$f({\underline z}, {\underline w})\in {\cal H}_0$ obey the partial 
differential 
equation in $(\ref{4.19})$. In both cases tracelessness leads to the UR being 
multiplicity free, apart from being complete in the sense that all 
$SU(3)$ UIR's do appear.  

\section{Concluding remarks}
To conclude, we have brought out the difficulties one encounters in 
naively  extending the Schwinger $SU(2)$ construction to $SU(3)$ 
particularly if one wishes to retain the simplicity and economy 
intrinsic to the $SU(2)$ case. We have shown how these difficulties 
can be overcome by exploiting the group $Sp(2, R)$ to obtain a 
`Generating Representation' of $SU(3)$ based on six bosonic oscillators. 
This $UR$ of $SU(3)$ contains all the representations of $SU(3)$ exactly
once. Further, we have shown how this `Generating Representation' for 
$SU(3)$ can also be constructed using the theory of induced representations 
and have constructively established the equivalence between the two by making 
use of the Bargmann representation. It is hoped that the construction 
presented here will have useful applications in various branches of physics 
much the same way as the $SU(2)$ construction has. Indeed, the work presented 
here has direct relevance to $SU(3)$ coherent states as will be shown in a
succeeding publication.   

\vskip1cm
\noindent
{\bf Acknowledgements}

This work was begun while NM was a Jawaharlal Nehru Chair Professor of
the University of Hyderabad. The hospitality extended to him during his stay
by the School of Physics, University of Hyderabad is gratefully acknowledged. 

\newpage
\def\theequation{A.\arabic{equation}}
\appendix{\bf{Appendix}: Boson operator construction of $SU(3) 
\times Sp(2,R)$ basis states}
\setcounter{equation}{0}

We give here the explicit construction of the orthonormal basis states
$|p,q; IMY; m>$ for ${\cal H}$ introduced in eqn.(\ref{3.26}).  We deal 
first with the states $|p,q; IIY;k>\in {\cal H}^{(p,q;0)}\subset
{\cal H}^{(p,q)}\bigcap {\cal H}_0$ having highest $SU(2)$ weight; then 
by repeated use of the $Sp(2,R)$ raising operator 
$K_+=\underline{{\hat a}}^{\dag}\cdot \underline{{\hat b}}
^{\dag}$ with $|p,q; IIY; m>\in{\cal H}^{(p,q;m-k)}\subset
{\cal H}^{(p,q)}$; and finally with the general state $|p,q;IMY;m>$
using the $SU(2)$ lowering operator.  At each stage the normalisation 
will be ensured. 

     As is well known, the boson operators 
${\hat a}^{\dag}_j, {\hat b}_j^{\dag}$ carry the following $U(2)$ 
quantum numbers\cite{10}:
     \begin{eqnarray}
     \matrix{
     &I\;\;\;&M\;\;\;&Y\cr
     {\hat a}^{\dag}_1,{\hat a}^{\dag}_2\;\;\;&1/2\;\;\;&\pm 1/2\;\;
     \;&1/3\cr 
     {\hat a}^{\dag}_3\;\;\;&0\;\;\;&0\;\;\;&-2/3\cr
     {\hat b}_2^{\dag},-{\hat b}_1^{\dag}\;\;\;&1/2\;\;\;&\pm 1/2\;\;\;&-1/3\cr
     {\hat b}^{\dag}_3\;\;\;&0\;\;\;&0\;\;\;&2/3}
     \end{eqnarray}

\noindent
Therefore, ${\hat a}^{\dag}_{\alpha} {\hat b}^{\dag}_{\alpha} 
\equiv {\hat a}^{\dag}_1
{\hat b}^{\dag}_1 +{\hat a}^{\dag}_2 {\hat b}^{\dag}_2$, 
${\hat a}^{\dag}_3$ and ${\hat b}_3^{\dag}$ are $SU(2)$ scalars.  
The I-Y multiplets present in the $SU(3)$ UIR $(p,q)$ are listed in eqn.(\ref{2.5}), and are parametrised by
two integers $r,s$.  The state $|p,q; IIY;k> \in {\cal H}_0$ 
involves $p$ factors ${\hat a}^{\dag}$ and $q$  factors ${\hat b}^{\dag}$ 
acting on the Fock
vacuum $|\underline{0}, \underline{0}>$, and in addition it is
annihilated by $K_- = \underline{{\hat a}}\cdot \underline{{\hat b}}$.  We
therefore start with the expression (guided by (A.1)):
     \begin{eqnarray}
     |p,q; IIY; k> &=& \left({\hat a}_1^{\dag}\right)^r
     \left({\hat b}_2^{\dag}\right)^s \sum\limits^{(p-r,q-s)_{<}}
     _{n=0,1,\ldots} C_n \left({\hat a}_{\alpha}^{\dag} 
      {\hat b}_{\alpha}^{\dag}
     \right)^n \left({\hat a}_3^{\dag}\right)^{p-r-n}
     \left({\hat b}_3^{\dag}\right)^{q-s-n} |
     \underline{0}, \underline{0}>,\nonumber\\
     r = I+\frac{Y}{2}&+& \frac{1}{3}(p-q),\;s=I-\frac{Y}{2} +
     \frac{1}{3} (q-p) .
     \end{eqnarray}

\noindent
The condition
     \begin{eqnarray}
     K_- |p,q; \;IIY;\;k > = 0
     \end{eqnarray}

\noindent
gives the recursion relation
     \begin{eqnarray}
     n(r+s+n+1) C_n = -(p-r-n+1)(q-s-n+1) C_{n-1}, n=1,2,\ldots,
     \end{eqnarray}

\noindent
with the solution
     \begin{eqnarray}
     C_n = \frac{(-1)^n}{n!}\;\frac{(p-r)! (q-s)! (r+s+1)!}
     {(p-r-n)! (q-s-n)! (r+s+n+1)!}\;C_0, n=1,2,\ldots .
     \end{eqnarray}

\noindent
Using this in eqn.(A.2), and after some algebra, the normalised
state is found to be:
     \begin{eqnarray*}
     |p,q;\;IIY;\;k > = {\cal N}_{pqIY}
     \frac{\left({\hat a}^{\dag}_1\right)^r}{r!} \frac{\left({\hat b}^{\dag}
     _2\right)^s}{s!} \;\times
     \end{eqnarray*}
     \begin{eqnarray*}
     \sum\limits^{(p-r,q-s)_<}_{n=0,1,\ldots}\;
     \frac{(-1)^n}{(r+s+n+1)!}\;
     \frac{\left({\hat a}^{\dag}_{\alpha} {\hat b}^{\dag}_{\alpha}\right)^n}
     {n!}\;
     \frac{\left({\hat a}_3^{\dag}\right)^{p-r-n}}{(p-r-n)!}\;
     \frac{\left({\hat b}_3^{\dag}\right)^{q-s-n}}{(q-s-n)!}\;
     |\underline{0},\underline{0}>\in{\cal H}^{(p,q;0)} ,
     \end{eqnarray*}
     \begin{eqnarray}
     {\cal N}_{pqIY} = \left\{r! s! (r+s+1)! (p-r)! (q-s)! 
     (p+s+1)! (q+r+1)!\big/ (p+q+1)!\right\}^{1/2} .
     \end{eqnarray} 

From eqn.(\ref{3.27b}) we know that vectors in ${\cal H}^{(p,q; m-k)}$ for 
$m>k$ are obtained from vectors in ${\cal H}^{(p,q;0)}$ by applying 
$K_+^{m-k}$.  Further, the normalisation is controlled by eqn. (3.25b).
We thus obtain:
     \begin{eqnarray}
     |p,q; \;IIY;\;m> =\left\{
     (2k-1)!\big/(m-k)! (m+k-1)!\right\}^{1/2}
     \left(\underline{{\hat a}}^{\dag}\cdot \underline{{\hat b}}^{\dag}
     \right)^{m-k}\;&&|p,q;\;IIY;\;k>\nonumber\\
     &&\in {\cal H}^{(p,q; m-k)} .
     \end{eqnarray}

The last step is to reach a general value $M\leq I$ for the
$SU(2)$ magnetic quantum number.  For this we apply the $SU(2)$ 
lowering operator $J_-={\hat a}^{\dag}_2 {\hat a}_1 - {\hat b}^{\dag}_1 
{\hat b}_2 \;(I-M)$ times to the state (A.7), keeping track of
normalisation.  This leads to the result:
     \begin{eqnarray}
     |p,q;\;IMY;\;m> = \left\{(I+M)!\big/
     2I! (I-M)!\right\}^{1/2}
     \left({\hat a}^{\dag}_2 {\hat a}_1 -{\hat b}_1^{\dag} 
     {\hat b}_2\right)^{I-M}|p,q;\; IIY;\;m > .
     \end{eqnarray}

\noindent
If we combine eqns.(A.6,7,8) we get the complete expression
     \begin{eqnarray*}
     |p,q;\;IMY;\;m> = {\cal N}_{pqIY} \left\{(2k-1)!
     (I+M)! (I-M)!\big/(m-k)!(m+k-1)! 2I!\right\}^{1/2}
     \times
     \end{eqnarray*}
     \begin{eqnarray*}
     (\underline{{\hat a}}^{\dag}\cdot \underline{{\hat b}}^{\dag})^{m-k}\;
     \sum\limits^{I-M}_{L=0}\;
     \sum\limits^{(p-r, q-s)_<}_{n=0}\;
     \frac{(-1)^{n+I-M-L}}{(r+s+n+1)!}\cdot
     \frac{\left({\hat a}^{\dag}_{\alpha}{\hat b}^{\dag}_{\alpha}\right)^n} 
      {n!}\times 
     \end{eqnarray*}
     \begin{eqnarray}
     \frac{\left({\hat a}_3^{\dag}\right)^{p-r-n}}{(p-r-n)!}\;
     \frac{\left({\hat b}_3^{\dag}\right)^{q-s-n}}{(q-s-n)!}\;
     \frac{\left({\hat a}^{\dag}_1\right)^{r-L}}{(r-L)!}\;
     \frac{\left({\hat a}_2^{\dag}\right)^L}{L!}\;
     \frac{\left({\hat b}_2^{\dag}\right)^{s-I+M+L}}{(s-I+M+L)!}\;
     \frac{\left({\hat b}_1^{\dag}\right)^{I-M-L}}{(I-M-L)!}\;
     |\underline{0},\underline{0}>.
     \end{eqnarray}    
 
We thus have explicit expressions for all the normalised basis states 
$|p,q; IMY; m >$ of ${\cal H}$.


\newpage
\begin{references}
\bibitem{1} 
J. Schwinger, {\it On angular momentum}, USAEC Report NYO-3071 (1952) 
(unpublished); reprinted in {\it Quantum theory of angular momentum}, L. C. 
Biedenharn and H. van Dam (eds), Academic Press, New York (1965); also in 
{\it A quantum legacy - Seminal papers of Julian Schwinger}, 
Kimball A. Milton (ed), World Scientific Publishing Company, Singapore (2000).
\bibitem{2} D. P. Arovas and A. Auerbach, Phys. Rev. B {\bf 38}, 316 (1988); 
A. Auerbach and D. P. Arovas, Phys. Rev. Lett. {\bf 61}, 617 (1988); 
A. Auerbach, {\it Interacting electrons and quantum magnetism}, (Springer, 
New York, 1994). 
\bibitem{3} 
Arvind, B. Dutta, N. Mukunda and R. Simon, Phys. Rev. A {\bf 52}, 
1609 (1993). 
\bibitem{4} K. Sundar, N. Mukunda and R. Simon, J. Opt. Soc. Am. A {\bf 12}, 
560 (1995). 
\bibitem{5} M. V. Berry and J. M. Robbins,  Proc. Roy. Soc. Lond. A {\bf 
453}, 1771 (1997). 
\bibitem{6} V. Bargmann, Rev. Mod. Phys. {\bf 34}, 829 (1962).
\bibitem{7} N. Mukunda and B. Radhakrishnan, J. Math. Phys. {\bf 15}, 1320, 
1332, 164, 1656 (1974). 
\bibitem{8} M. Moshinsky, Rev. Mod. Phys. {\bf 34}, 813 (1962).
\bibitem{9} G. W. Mackey, {\it Group representations in Hilbert Space}, 
(AMS, Providence, Rhode Island, 1963).
\bibitem{10} See, for instance, J. J. de Swart, Rev. Mod. Phys. {\bf 35}, 
916 (1963). 
\bibitem{11} N. Mukunda and L. K. Pandit, J. Math. Phys. {\bf 6}, 746 (1965). 
\bibitem{12}  D. Sen and M. Mathur, J. Math. Phys. {\bf 42}, 4181 (2001). 
 \bibitem{13} N. Mukunda and L. K. Pandit Prog. Theor. Phys. {\bf 34}, 46 
(1965). 
\bibitem{14} See, for instance, A. M. Perelomov, Usp. Fiz. Nauk. {\bf 123}, 
23 (1977) [ Sov. Phys. Usp. {\bf 20}, 703 (1977)]; K. W\'odkiewicz and J. H. 
Eberly J. Opt. Soc. Am. B {\bf 2}, 458 (1985).
\bibitem{15} V. Bargmann, Ann. Math. {\bf 48}, 568 (1947). 
\bibitem{16} V. Bargmann, Commun. Pure Appl. Math. {\bf 14}, 187 (1961). 

\end{references}
\end{document}